\documentclass[aip,graphicx,nofootinbib]{revtex4-1}

\draft 

\usepackage{cancel}
\newcommand{\p}{\partial}
\newcommand{\nn}{\nonumber}
\renewcommand{\thefootnote}{\alph{footnote}}
\usepackage{amsfonts}
\usepackage{hyperref}

\usepackage{fancyhdr}

\begin{document}


(Submitted to JMP )
\title{Magnetic monopoles in noncommutative quantum mechanics} 



\author{Samuel Kov\'a\v{c}ik}
\affiliation{School of Theoretical Physics,
Dublin Institute for Advanced Studies,
10 Burlington Road, Dublin 4, Ireland.}
 \email{skovacik@stp.dias.ie}
 \altaffiliation[previously at ]{2)}
\author{Peter Pre\v{s}najder}
\affiliation{Faculty of Mathematics, Physics and Informatics,
Comenius University Bratislava, Mlynsk\'a dolina, 
Bratislava, 842 48, Slovakia.}
\email{presnajder@fmph.uniba.sk}

\date{\today}

\begin{abstract}
We discuss  certain generalization of the Hilbert space of states in noncommutaive quantum mechanics that, as we show, introduces magnetic monopoles into the theory. Such generalization arises very naturally in the considered model, but can be easily reproduced in ordinary quantum mechanics as well. This approach offers a different viewpoint on the Dirac quantization condition and other important relations for magnetic monopoles. We focus mostly on the kinematic structure of the theory, but investigate also a dynamical problem (with the Coulomb potential).
\end{abstract}

\small{\ Ref. number: DIAS-STP-16-11}

\maketitle 


\section{Introduction}

Magnetic monopoles are a unique part of physics. Their existence is being considered for more than a century, yet they have never been observed. They appear (in theory) in various areas of physics, persistently throughout different models, always playing a slightly different role.

They premiered in the classical theory of electromagnetism. Maxwell equations in vacuum are symmetric under a transformation known as electric-magnetic duality $\left(\textbf{E},\textbf{B}\right) \rightarrow \left( \textbf{B}, -\textbf{E}\right)$. This symmetry is violated in the presence of electric sources $\rho_E$, but can be recovered by introducing (monopole) magnetic sources $\rho_M$. 

New phenomena appear in such a generalized theory, for example, electromagnetic fields generated by a static system of electric and magnetic monopole have a non-vanishing angular momentum. 

It is often comfortable to work with electromagnetic potentials $\textbf{A}, \varphi$ instead of electromagnetic fields $\textbf{E}, \textbf{B}$. It might seem that magnetic potentials cannot describe magnetic monopoles, since $\mbox{div rot } \textbf{A} = 0$ seems to follow directly (as $ \partial_{[ i} \partial_{j ]} A_k=0$), resulting in the absence of magnetic monopoles. This, however, holds only for nonsingular potentials $\textbf{A}$ (for which the order of derivatives can be exchanged) and, therefore, monopoles could be described by singular potentials, see e.g. \cite{zwanziger, milton}. The following potentials
\begin{equation}
\textbf{A} = \frac{g}{4 \pi r} \frac{\textbf{r}\times \textbf{n}}{r-\textbf{r} . \textbf{n}}\ \ \ \ \mbox{or}\ \ \ \ \textbf{A} = - \frac{g}{4 \pi r} \frac{\textbf{r}\times \textbf{n}}{r+\textbf{r} . \textbf{n}}
\end{equation}
(where $\textbf{n}$ is a unit constant vector) result into Coulomb(-like) magnetic field
\begin{equation}
\textbf{B} = \frac{g}{4 \pi} \frac{\textbf{r}}{r^3}.
\end{equation}

In the quantum theory the description of Yang \cite{Yang} is preferred. In this framework, one describes monopoles with sections (avoiding the singularity) related by a gauge condition in the overlapping regions. A consistent quantum theory requires the electric and the magnetic charge to satisfy the Dirac quantization condition (in convenient units)
\begin{equation}
e g  = \frac{n}{2} \, , n \in \mathbb{Z} \,.
\end{equation}
This condition has an appealing physical consequence - the electric charge has to be quantized, as is observed in nature.

The appearance of magnetic monopoles in quantum field theory is (again) slightly different, they appear as topological solutions, in contrast to ordinary particles appearing as quantum excitations. As was shown by Polyakov and 't Hooft,  monopoles are a general consequence of grand unification theories (GUT), appearing when a higher symmetry brakes down into a product containing $U(1)$, \cite{pol, hooft}. Mass of the monopoles is, therefore, expected to be on the GUT breaking scale. Monopoles also appear in cosmology, existing as topological defects between domains of different vacua. Cosmology also offers an explanation why we have not observed any magnetic monopoles yet, the process of inflation diluted them remarkably.

For the purpose of this paper is the quantum mechanical (QM) description the most convenient one. Let us quote the results of Zwanziger \cite{zwanziger}, in the presence of monopole states is the usual Heisenberg algebra modified as
\begin{eqnarray} \label{MMzw1}
\left[ \hat{x}_i , \hat{x}_j \right]  &=& 0 ,  \\ \nn
\left[ \hat{\pi}_i , \hat{x} _j \right]  &=& - i \delta_{ij} , \\ \nn
\left[\hat{\pi}_i ,\hat{\pi}_j  \right]  &=& i \mu \varepsilon_{ijk} \frac{\hat{x}_k}{r^3}  ,
\end{eqnarray}
where $\mu = e g$ and the Dirac quantization condition dictates $\mu \in \mathbb{Z} /2$. 

In the same paper, a dynamical problem with the Coulomb potential was analyzed. The Coulomb problem in ordinary QM can be solved algebraically, as was first proposed by Pauli, generalizing the classical notion of the Laplace-Runge-Lenz vector $A_i$, \footnote{Not to be confused with the electromagnetic potential.}. Components of this vector, together with the components of the angular momentum operator form a representation of either the $so(1,3)$ or the $so(4)$ algebra, depending on the sign of the energy of the system. The algebra closes only on energy eigenstates $\hat{H}\psi = E \psi$

\begin{eqnarray} \label{LRLalg}
&\left[ \hat{L}_i, \hat{L}_j \right] =&i \varepsilon_{ijk} \hat{L}^k ,\\   \nn
&\left[ \hat{L}_i, \hat{A}_j \right] =& i\varepsilon_{ijk} \hat{A}^k  ,\\  \nn
&\left[ \hat{A}_i, \hat{A}_j \right] =& -  2i  \varepsilon_{ijk} \hat{H} \hat{L}^k,\\ \nn
&\left[ \hat{L}_i, \hat{H} \right] =&\left[ \hat{A}_i, \hat{H} \right]= 0  .
\end{eqnarray}
These relations are indifferent to the presence of monopole states, however, the Casimir operators, which determine the energy spectrum, are not
\begin{eqnarray} \label{Zwanziger}
\hat{A}_i \hat{L}_i &=& q \mu  ,  \\ \nn
\hat{A}_i \hat{A}_i -   2 \hat{H} \left( \hat{L}_i \hat{L}_i +1\right) &=& q^2+ 2 \hat{H} \left(- \mu^2  \right) ,
\end{eqnarray}
where $q$ is the electric charge from the Coulomb potential.

Below we shall investigate magnetic monopoles in the framework of noncommutative quantum mechanic (NC QM), which is a particular application of the ideas of noncommutative geometry to QM, \cite{Con, Mad1}. NC QM differs from ordinary QM by having a nonvanishing commutator of the coordinate operators. This results in the impossibility of exact position measurements, which can be motivated by (thought experiments in) quantum theory of gravity, \cite{DFR}. NC theories are closely related to different candidates for such a theory, the string/M-theory being a prominent example, \cite{string}.

NC QM models that do not possess rotational invariance have been investigated in \cite{AB}, \cite{CP}, however, our problem requires full 3D rotational symmetry. Such a model was proposed in \cite{Jabbari}, the construction used here has been developed in \cite{GP1,GP2,vel,LRL}. Using the auxiliary bosonic operators approach, the exact solution of NC Coulomb problem was found, both dynamically and algebraically. In this paper, we utilize the same approach, but consider a generalized class of physical states to describe magnetic monopoles.

This paper is organized as follows. First, we construct NC QM using auxiliary bosonic operators. In subsection A we analyze general kinematical structures, in subsection B a dynamical one (with the Coulomb potential). In subsection C we briefly present how can the results be reproduced in the context of ordinary QM. Conclusions are followed by the Appendix containing all lengthy and technical calculations.

\section{Noncommutative quantum mechanics}
The first thing to do is to decide on the RHS of the noncommutativity relation, from which the restriction on position measurements follows. We study a rotationally invariant model described by
\begin{equation} \label{NCrel}
[x_i, x_j] = 2 i \lambda \varepsilon_{ijk} x_k  ,
\end{equation}
where $\varepsilon_{ijk}$ is the Levi-Civita symbol and $\lambda$ is a constant describing the NC length scale. It is not fixed, but as an artifact of quantum gravity it could be expected to be approximately the Planck length. Resulting NC space corresponds to an infinite sequence of fuzzy spheres.

Let us consider two set of auxiliary bosonic creation and annihilation (c/a) operators satisfying
\begin{equation}\label{aux}
[a_\alpha,a^+_\beta]=\delta_{\alpha\beta },\ \
[a_\alpha,a_\beta]=[a^+_\alpha, a^+_\beta]=0,
\end{equation}
with $ \alpha , \beta = 1,2$, which act in a Fock space $\mathcal{F}$ spanned on normalized vectors
\begin{equation}
|n_1,n_2\rangle= \frac{(a^+_1)^{n_1}\,(a^+_2)^{n_2}}{
\sqrt{n_1!\,n_2!}}\ |0\rangle.
\end{equation}
The NC coordinates satisfying (\ref{NCrel}) are constructed using the c/a operators as
\begin{equation}\label{NCx}
x_i = \lambda \sigma^i_{\alpha \beta} a^+_\alpha a_\beta  ,
\end{equation}
where $\sigma^i$ are the Pauli matrices. Using the number operator $N=a^+_\alpha a_\alpha$ we can define the radial coordinate operator as
\begin{equation}
r = \lambda \left( N+1 \right) = \lambda \left( a^+_\alpha a_\alpha +1 \right)  .
\end{equation}

It can be easily checked that $r^2  = x^2 + \lambda^2$, which differs from the ordinary result but reproduces it in the $\lambda \rightarrow 0$ limit. We refer to such as 'the commutative limit', since in it (\ref{NCrel}) becomes $[x_i,x_j]=0$ as in ordinary QM. In this limit should the results either reproduce the ordinary ones or vanish.

We define the Hilbert space $\mathcal{H}_\kappa $ as a completion of the linear space of operators in the auxiliary Fock space spanned by analytic functions $\Psi_\kappa(a^+,a)$ satisfying relation 
\begin{equation} \label{states}
\Psi_\kappa(e^{-i\tau} a^+,e^{i\tau} a) = e^{-i\tau\kappa} \Psi_\kappa (a^+,a), \ \tau \in {\bf R}, \  \mbox{fixed}\  \kappa \in \mathbb{Z}  ,
\end{equation}
that possesses finite norm
\begin{equation} \label{norm}
||\psi_\kappa||^2 = 4\pi \lambda ^2 Tr [ \psi^+_\kappa \,\hat{r}\, \psi_\kappa] ,
\end{equation} 
where $\hat{r}$ acts as $\hat{r} \psi_\kappa = \frac{1}{2}\left( r \psi_\kappa + \psi_\kappa r \right)$ and has been added to reproduce the ordinary integration $\int d^3x$ in the commutative limit \footnote{This can be checked by computing the volume of a ball with radius $R \gg \lambda$.}
Because NC coordinates (\ref{NCx}) contain equal number of creation and annihilation operators, for any state of the form $\Psi_0(\textbf{x})$ is $\kappa=0$. In this paper we consider a generalized class of states with $\kappa\neq 0$. If $\kappa < 0$ there is $|\kappa|$ more annihilation than creation operators $\# a - \# a^+ = -\kappa = |\kappa|$, if $\kappa > 0 $ it is vice versa $\#a^+ - \#a =  \kappa$.
\subsection{Kinematic structures}
We shall now define important physical operators on $\mathcal{H}_\kappa$. To distinguish them from the ones on the auxiliary space $\mathcal{F}$, we denote them with a hat. We are using a lower index to distinguish between left and right multiplication
\begin{eqnarray}
\hat{X}_{i,L} \Psi_\kappa = x_i \Psi_\kappa ,\ &   & \hat{X}_{i,R} \Psi_\kappa  = \Psi_\kappa x_i ,\\ \nonumber
\hat{r}_{L} \Psi_\kappa = r \Psi_\kappa , \ &  & \hat{r}_{R} = \Psi_\kappa r  .
\end{eqnarray}
The operators $\hat{X}_{i,L}$ and $\hat{X}_{i,R}$ carry the $so_L(3)$ and the $so_R(3)$ Lie algebra representation respectively. Coordinate operators on $\mathcal{H}_\kappa$ are defined as symmetrical combinations
\begin{equation} \label{Xnc}
\hat{X}_i = \frac{1}{2} \left( \hat{X}_{i,L} + \hat{X}_{i,R}\right) , \ \hat{r} =  \frac{1}{2} \left( \hat{r}_{L} + \hat{r}_{R}\right),  
\end{equation}
while the angular momentum operator satisfying $[\hat{L}_i, \hat{L}_j] = \varepsilon_{ijk} \hat{L}_k$ is an antisymmetrical one
\begin{equation}
\hat{L}_i = \frac{1}{2\lambda} \left( \hat{X}_{i,L} - \hat{X}_{i,R}\right).
\end{equation}
Note that the angular momentum operator acts on $\psi_\kappa$ as $\hat{L}_i \psi_\kappa = \frac{1}{2\lambda} [x_i , \psi_\kappa]$ and that \footnote{This follows from the fact that the right multiplication changes the order in commutator, generating an extra minus sign and that $\hat{L}_k \propto \hat{X}_{k,L} - \hat{X}_{k,R}$.}
\begin{equation} \label{[x,x]}
[\hat{X}_i, \hat{X}_j] = \lambda^2 \varepsilon_{ijk}\hat{L}_k.
\end{equation}
Even thought for generalized states $\hat{r}_L \neq \hat{r}_R \neq \hat{r}$, they are closely related for each $\mathcal{H}_\kappa, \ \kappa \in \mathbb{Z}$
\begin{eqnarray}
\hat{r}&=&\hat{r}_L - \frac{\lambda \kappa}{2} = \hat{r}_R + \frac{\lambda \kappa}{2}, \\ \nn
 \lambda \kappa &=& \hat{r}_L- \hat{r}_R.
\end{eqnarray}
As $\hat{r}$ commutes with the generators $\hat{X}_{i,L}$ and $\hat{X}_{i,R}$, the $so_L(3)$ and the $so_R(3)$ Casimir operators can be expressed in terms of $\hat{r}$ and $\kappa$ as 
\begin{eqnarray}
\hat{X}^2_L &=& \left( \hat{r}+ \frac{\lambda \kappa}{2} \right)^2 - \lambda^2, \\ \nn
\hat{X}^2_R &=& \left( \hat{r}- \frac{\lambda \kappa}{2} \right)^2- \lambda^2 .
\end{eqnarray}
For states $\psi_0(\textbf{x})$ with $\kappa=0$ it holds that $\hat{r}_L = \hat{r}_R$ and $\hat{r} = \hat{r}_L$ can be chosen for simplicity, as was done in the aforementioned references. We can use their definitions and results, sharing the same line of reasoning, but have to replace $\hat{r}_L \rightarrow \hat{r}$ and check for possible consequences and modifications.

The functions $\psi_\kappa$ with fixed $\kappa$ are mappings $\mathcal{F}_n \rightarrow \mathcal{F}_{n+\kappa}$, they form a representational space for an irreducible $SO(4)$ representation in which it holds that $\hat{r} = \lambda \left( n+1 \right) + \frac{\lambda \kappa}{2}$. The Casimir operators are \footnote{Note that by eliminating $\hat{r}$ they can be combined into a single equation  $2 \hat{c}_1 = \left( \frac{2}{\kappa}\right)^2 \hat{c}^2_2 + \left( \frac{\kappa}{2}\right)^2 -1$.}
\begin{eqnarray}\label{cas4} 
\hat{c}_1 &=&\hat{L}^2 + \frac{1}{ \lambda^2} \hat{X}^2 = \frac{1}{4 \lambda^2} \left( \hat{X}^2_L + \hat{X}^2_R \right) = \frac{1}{2\lambda^2}\left( \hat{r}^2-\lambda^2+\left( \frac{\lambda \kappa}{2} \right)^2 \right), \\  \nn
\hat{c}_2 &=&\frac{1}{2\lambda} \hat{X}_i\hat{L}_i =\frac{1}{4\lambda^2} \left( \hat{X}^2_L - \hat{X}^2_R \right) = \frac{\kappa}{2\lambda}\hat{r},
\end{eqnarray}

Two of the most important physical operators, namely the free Hamiltonian and the velocity operators are defined as
\begin{eqnarray} 
\hat{H}_0 \psi_\kappa &=& \frac{1}{2\lambda \hat{r}} [ a^+_\alpha , [ a_\alpha , \psi_\kappa ]] , \\ \nn
\hat{V}_i \psi_\kappa &=&i [ \hat{H}_0,\hat{X}_i]\psi_\kappa= \frac{i}{2 \hat{r}} \sigma^i_{\alpha \beta} \left(a^+_\alpha \psi_\kappa  a_\beta - a_\beta \psi_\kappa  a^+_\alpha \right)  .
\end{eqnarray}
To begin revealing the overall structure let us first combine $\hat{X}_i$ and $\hat{L}_i$ together as
\begin{equation}
\hat{L}_{ij} = \varepsilon_{ijk} \hat{L}_k , \ \hat{L}_{k4} = - \hat{L}_{4k} = \lambda^{-1} \hat{X}_k \,
\end{equation}
to observe an $so(4) \cong su_L(2) \oplus su_R(2)$ Lie algebra structure
\begin{equation}
[\hat{L}_{ab},\hat{L}_{cd}] = i \left( \delta_{ac} \hat{L}_{bd}-\delta_{bc} \hat{L}_{ad}-\delta_{ad} \hat{L}_{bc} + \delta_{bd} \hat{L}_{ac} \right) ,
\end{equation}
where indices go over as $i,j,k,...=1,2,3$ and $a,b,c, ... = 1,...,4$. 

The central point of ordinary QM is the Heisenberg uncertainty relation, the commutator of $[\hat{V}_i, \hat{X}_j]$. In \cite{vel} it has been shown that this relation obtains a $\lambda$-correction already for $\psi_0$ states and as it turns out, this correction is of the same for $\psi_\kappa$ states as well
\begin{equation} \label{xV}
[\hat{X}_i,\hat{V}_j] = i \delta_{ij} \left(1-\lambda^2 \hat{H}_0\right) \equiv i \lambda \delta^{ij} \hat{V}_4  ,
\end{equation}
where $\hat{V}_4 \psi_\kappa = \frac{1}{\lambda}- \lambda \hat{H}_0= \frac{1}{2 \hat{r}}\left( a^+_\alpha \psi_\kappa a_\alpha + a_\alpha \psi_\kappa a_\alpha^+ \right)$. Note that $\hat{V}_a$ transforms as an $SO(4)$ vector. 

Another interesting result of \cite{vel} is that even though the coordinates do not commute, the velocities do. This, however, fails to be true for $\kappa \neq 0$ states, instead it holds
\begin{equation} \label{F}
\left[\hat{V}_i,\hat{V}_j \right] = i\hat{F}_{ij} ,
\end{equation}
with the magnetic field strength given as 
\begin{equation}
\hat{F}_{ij} = \varepsilon_{ijk} \frac{- \frac{\kappa}{2}\hat{X}_k}{\hat{r}(\hat{r}^2-\lambda^2)}.
\end{equation}
This can be generalized into an $SO(4)$ structure by noting that
\begin{equation}
\hat{F}_{ab} =-i[\hat{V}_a,\hat{V}_b]= -\frac{\kappa \lambda}{2}\frac{ \varepsilon_{abcd} \hat{L}_{cd}}{\hat{r}\left(\hat{r}^2-\lambda^2\right)}. 
\end{equation}
For its square it holds that
\begin{equation} \label{F42}
\frac{1}{2}\hat{F}_{ab}^2=\frac{1}{2}\hat{F}_{ab} \hat{F}_{ab} =\frac{1}{2} \left( \frac{\kappa }{2}\right)^2 \frac{\left(r^2 - \lambda^2 +\left( \frac{\kappa}{2}\right)^2\right)}{r^2 \left( r^2 - \lambda^2 \right)^2} \, .
\end{equation}

In the aforementioned reference it was noted that the eigenvalues of $\hat{V}_a^2$ lay on a $S^3$ sphere with a radius of $\lambda^{-1}$. This structure is modified for $\kappa \neq 0$ states as well
\begin{equation} \label{V42}
\hat{V}_a^2 = \frac{1}{\lambda^2} \left( 1 - \frac{\left( \frac{\kappa \lambda}{2}\right)^2}{\hat{r}^2 - \lambda^2}\right) .
\end{equation}
In the classical theory such a ($\kappa$ dependent) term arises due to the effective potential of the angular momentum of the fields. Note the similar terms appearing on the RHS of (\ref{F42}, \ref{V42}) and the equations for Casimir operators (\ref{cas4}), they allow us to express the squares as
\begin{equation}
\hat{V}_a^2 = \frac{1}{\lambda^2} - \frac{\lambda^2 \hat{c}^2_2}{\hat{r}^2 \left( \hat{r}^2 - \lambda^2\right)} , \ \hat{F}^2_{ab} = \frac{\lambda^4}{\hat{r}^4 \left(\hat{r}^2 - \lambda^2\right)^2}\hat{c}_1 \hat{c}_2^2 .
\end{equation}
We can combine these equations to obtain a single one, generalizing the important $\kappa=0$ result $\hat{V}_a^2 = \lambda^{-2}$ to 
\begin{equation}
\hat{V}_a^2 + \hat{\varphi}\hat{F}_{ab}^2 = \lambda^{-2} , \,\hat{\varphi}= \frac{\hat{r}^2 \left( \hat{r}^2 - \lambda^2 \right)}{\hat{r}^2 -\lambda^2 + \left(\frac{\lambda \kappa}{2}\right)^2}.
\end{equation}
%
%
\subsection{Dynamical structure}
Before drawing any conclusion let us take a look at a certain dynamical structure. It is convenient to choose the Coulomb potential $U=\frac{q}{r}, \ q = e^2$, there are two reasons for it. First, the Coulomb problem can be solved algebraically (as was found by Pauli) and second, it has already been analyzed in the framework of NC QM (for $\kappa=0$ states) \cite{GP1,GP2,LRL} .

The time independent Schr\"odinger equation with the Coulomb potential for generalized $\kappa$ states is
\begin{equation} \label{SchrK}
\hat{H}\psi_\kappa = \left( \hat{H}_0 - \frac{q}{\hat{r}}\right) \psi_{\kappa} = E \psi_{\kappa} .
\end{equation}

The following vector is called the Laplace-Runge-Lenz (LRL) vector\footnote{Even thought it was in fact first discovered by Jakob Hermann and Johann Bernoulli.} and is conserved in the presence of such a potential

\begin{equation}
\hat{A}_k = \frac{1}{2} \varepsilon_{ijk} \left( \hat{L}_i \hat{V}_j + \hat{V}_j \hat{L}_i \right) + q \frac{\hat{X}_k}{\hat{r}} .
\end{equation}
The same is true for the angular momentum operator. We can express it as
\begin{equation}
[\hat{H} , \hat{L}_i ] = 0 , \, [\hat{H}, \hat{A}_i] = 0  .
\end{equation}
Commutators of the angular momentum and the LRL vector are
\begin{eqnarray} \label{LRLalgNC}
[\hat{L}_i, \hat{L}_j ] &=& i \varepsilon_{ijk} \hat{L}_k  , \\ \nonumber 
[\hat{L}_i, \hat{A}_j] &=& i \varepsilon_{ijk} \hat{A}_k  , \\ \nonumber
[\hat{A}_i, \hat{A}_j] &=& -2i\hat{H} \left(1 - \lambda^2 \hat{H}\right) \varepsilon_{ijk} \hat{L}_k,
\end{eqnarray}
Restricting to energy eigenstates we can take $\hat{H} =E$ and obtain either the $so(3,1)$ or the $so(4)$ Lie algebra, depending on the sign of $E\left(1-\lambda^2 E\right)$. Following from the group theory we know that their Casimir operators are allowed to take discrete values only, from which the discreteness of the spectrum follows. For generalized $\kappa$ states the Casimir operators are
\begin{eqnarray} \label{cas}
\hat{C}_1 &=& \hat{L}_i \hat{A}_i = -\frac{\kappa}{2}q , \\ \nonumber 
\hat{C}_2 &=& \hat{A}_i \hat{A}_i + (- 2E +\lambda^2 E^2) (\hat{L}_i \hat{L}_i + 1)  \\ \nn
&=& q^2 + \left(\frac{\kappa}{2}\right)^2 (- 2E +\lambda^2 E^2) .
\end{eqnarray}
Again, we observe a $\kappa$ correction.

\subsection{Ordinary space}

It has been noted earlier that the results can be reproduced in ordinary QM. The starting point is to realize that the isometry group of three-dimensional Euclidean space is locally isomorphic to that of complex $\textbf{C}^2$ plane. Two complex coordinates $z_1, z_2$ of $\textbf{C}^2$ can be mapped into three real $\textbf{R}^3$ coordinates by (a Hopf fibration) $x_i = \bar{z} \sigma^i z$. This relation can be understood using Cayley-Klein parameters
\begin{eqnarray} \label{CP}
z_1 = \sqrt{r} \cos \left(\theta /2 \right) e^{\frac{i}{2}\left(\varphi+\gamma \right)}, \ && \bar{z}_1 = \sqrt{r} \cos \left(\theta /2\right) e^{-\frac{i}{2}\left(\varphi+\gamma \right)}, \\ \nn
z_2 = \sqrt{r} \sin \left(\theta /2 \right) e^{\frac{i}{2}\left(-\varphi+\gamma \right)}, \ &&\bar{z}_2 =\sqrt{r} \sin \left(\theta /2 \right) e^{-\frac{i}{2}\left(-\varphi+\gamma \right)} ,
\end{eqnarray}
which are by $x_i = \bar{z} \sigma^i z$ transformed into spherical coordinates of $\textbf{R}^3$, the angle $\gamma$ is lost in translation. 

$\textbf{C}^2$ is naturally equipped with a Poisson structure 
\begin{equation} \label{Poisson}
\{z_\alpha, \bar{z}_\beta\}=i \delta _{\alpha \beta}  .
\end{equation}
The (free) Hamiltoanian is 
\begin{equation} \label{cHam}
\hat{H}_0 = \frac{1}{2r} \{ \bar{z}_\alpha , \{ z_\alpha , . \} \} , \ H_0 \psi (z,\bar{z}) = -  \frac{1}{2r}\p_{\bar{z}_\alpha} \p_{z_\alpha} \psi (z,\bar{z}),
\end{equation}
where $r=\bar{z}_\alpha z_\alpha$. Using this we can define the velocity operator as
\begin{equation} \label{cV}
\hat{V}_i  =\{ \hat{X}_i, \hat{H}_0 \}= - \frac{i}{2r}\sigma^i_{\alpha \beta} (\bar{z}_\alpha \partial _{\bar{z}_\beta} + z_\beta \partial_{z_\alpha}),
\end{equation} 
the coordinate operator acting only as a left multiplication now.

Quantization of $\textbf{C}^n$ can be carried out by replacing $\bar{z}_\alpha , z_\alpha \rightarrow \sqrt{\lambda}a_\alpha^+, \sqrt{\lambda}a_\alpha$ and derivatives with commutators. Note that our model of NC QM can be reconstructed this way, for example the Hopf relation $x_i = \bar{z} \sigma^i z$ becomes (\ref{NCx}). 

If we restrict the algebra of functions to $\textbf{C}^2$ on only those of the form $\psi_0(\textbf{x})$, the Hamiltonian (\ref{cHam}) and velocity operator (\ref{cV}) are acting as in ordinary QM
\begin{equation}
\hat{H}_0 \psi_0(\textbf{x}) = - \frac{1}{2}\partial_i \partial_i \psi_0(\textbf{x}), \, \hat{V}_i \psi_0(\textbf{x}) = -i \partial_i \psi_0(\textbf{x})  ,
\end{equation}
as follows from the chain rule for derivatives. This way we can formulate ordinary QM on $\textbf{C}^2$ instead of $\textbf{R}^3$. 

We can also consider a generalized class of states
\begin{equation} \label{comStates}
\psi_\kappa (\textbf{x}, \xi) = \psi_0 (\textbf{x})  \xi , \ \xi = \sum \limits_\kappa{}^{'} C_{\kappa_1  \kappa_2} z_1^{\kappa _1}z_2^{\kappa _2} ,
\end{equation}
with the sum $ \sum \limits_\kappa{}^{'}$ going over all $\kappa_1, \kappa_2$ such that $\kappa_1 + \kappa_2 = - \kappa$. \footnote{Even more general $\xi=\sum  \limits_\kappa{}^{'} C_{\kappa_1 \kappa_2 \kappa_1' \kappa_2'} z_1^{\kappa_1}z_2^{\kappa_2}\bar{z}_1^{\kappa_1'}\bar{z}_2^{\kappa_2'}$ with $\kappa_1 + \kappa_2 - \kappa_1'-\kappa_2' = - \kappa$ could be used, but our choice simplifies the calculations and proves the same point.} This alters the action of (\ref{cV}) as
\begin{equation} \label{comVA}
\hat{V}^j \psi_\kappa= (- i \partial_j + \mathcal{A}^j)\psi_\kappa , \, \mathcal{A}_j = - \frac{i}{2r\xi} \sigma ^j _{\gamma \delta} z_\delta (\partial _{z_\gamma} \xi)  .
\end{equation} 
The gauge potential $\mathcal{A}_j$ satisfies (compare it with the last term in (\ref{V42}))
\begin{equation} \label{AA}
\frac{1}{2}(\mathcal{A}_j)^+\mathcal{A}_j = \left(\frac{\kappa}{2}\right)^2 \frac{1}{2r^2} ,
\end{equation}
The commutative limit of the results derived in NC QM can be obtained by considering states (\ref{comStates}), for example
\begin{equation} \label{VVcom}
[\hat{V}_i, \hat{V}_j] = -\frac{\kappa}{2}i\varepsilon^{ijk} \frac{\hat{X}^k}{r^3}.
\end{equation}

We are now ready to draw conclusions about our results and their relation to magnetic monopoles. 

\section{Summary and conclusions}

Let us recall the kinematic structure of ordinary QM in the presence of monopole states as was derived in \cite{zwanziger} (on the left) and compare it with the kinematic structure of NC QM with generalized $\kappa \neq 0$ states (equations (\ref{[x,x]}), (\ref{xV}), (\ref{F}) on the right)

\begin{equation} 
\begin{array}{lcl}
\ \left[ \hat{x}_i , \hat{x}_j \right]  = 0   & \leftrightarrow &  \ [\hat{X}_i, \hat{X}_j ] = \lambda^2 \varepsilon_{ijk} \hat{L}_k,   \\ 
\ \left[ \hat{x}_i, \hat{\pi}_j  \right]  =  i \delta_{ij} &\leftrightarrow & \  [\hat{X}^i,\hat{V}^j] = i \delta^{ij} \left(1-\lambda^2 \hat{H}_0\right), \\ 
\ \left[\hat{\pi}_i ,\hat{\pi}_j  \right]  = i \mu \varepsilon_{ijk} \frac{\hat{x}_k}{r^3}  & \leftrightarrow & \ \left[\hat{V}_i,\hat{V}_j \right]  = i\frac{-\kappa}{2} \varepsilon_{ijk} \frac{\hat{X}_k}{\hat{r}(\hat{r}^2-\lambda^2)}. 
\end{array} 
\end{equation}

The relations between the angular momentum operators and the LRL vector are the same (in the $\lambda \rightarrow 0$ limit), as one can check comparing (\ref{LRLalg}) and (\ref{LRLalgNC}).  Zwanziger \cite{zwanziger} derived the Casimir operators for the symmetry algebra of the Coulomb problem in the presence of magnetic monopoles (on the left). Let us compare his results with those for $\kappa$ states in equation (\ref{cas}) (on the right)
\begin{equation}
\begin{array}{lcl}
\hat{C}_1 = -q \mu  & \leftrightarrow &  \hat{C}_1= \frac{\kappa}{2}q , \\
\hat{C}_2 = q^2 + (\mu)^2(-2E) & \leftrightarrow & \ \hat{C}_2 = q^2 +
\left(\frac{\kappa}{2}\right)^2 (-2E+\lambda^2 E^2).
\end{array}
\end{equation}
The results are the same (in the commutative limit) if we set $\mu = -\frac{\kappa}{2}$. We need to check if such identification is possible, since $\mu$ has to obey the Dirac quantization condition $\mu \in \mathbb{Z}/2$. Recall that $\kappa$ counts the difference in the number of creation and annihilation operators and therefore $\kappa/2 \in\mathbb{Z}/2$ as well. The identification is perfect and offers a different viewpoint on the Dirac condition. Therefore $\psi_\kappa$ are to be interpreted as monopole states in NC QM. 

If we set $\lambda=0$, but keep $\kappa \neq 0$ we obtain ordinary QM with magnetic monopoles. By setting $\kappa = 0,\ \lambda \neq 0$ we obtain NC QM without monopoles. Finally by setting $\kappa = \lambda = 0$ ordinary QM (without monopoles) is recovered.

It shall be reminded that for a system of two dyons\footnote{Dyon is a particle with both the electric $e$ and the magnetic charge $g$.} are the parameters $q, \mu$ defined (in convenient units) as
\begin{equation}
q = - \frac{e_1 e_2 + g_1 g_2}{4\pi} , \, \mu =\frac{e_1 g_2 - g_1 e_2}{4\pi} \, .
\end{equation}
Therefore, the considered case describes for example an electron orbiting a nucleus with a magnetic monopole in it or an electrically charged magnetic monopole.

From (\ref{comStates}), it can be understood how do the generalized states describe monopoles. In $\textbf{C}^2$ there are 4 coordinates, but for wavefunctions of the form $\psi_0(\textbf{x})$ one of them, with a topology of $S^1$, vanishes. However, for $\psi_\kappa$ states it persists as a factor $e^{- \frac{i}{2}\kappa \gamma}$ winding around
\begin{eqnarray} 
\psi_{0}&=&\psi_0(\textbf{x})=\Phi(r,\varphi, \theta), \\ \nonumber
\psi_{1}&=&\psi_0(\textbf{x})\bar{z}_1=\Phi(r,\varphi, \theta) e^{-\frac{i}{2}\gamma}, \\ \nonumber
\psi_{2}&=&\psi_0(\textbf{x})\bar{z}_1\bar{z}_2 =\Phi(r,\varphi, \theta) e^{- i \gamma}, \\ \nn
&...& \\ \nn
\psi_\kappa &=& \Phi(r,\varphi, \theta) e^{- i \frac{\kappa}{2}\gamma}.
\end{eqnarray}

Note that $|\psi_\kappa|^2 = \psi_\kappa^\dagger \psi_\kappa$ always contains equal number of creation and annihilation operators (or $\bar{z}$ and $z$).

\section{Appendix}

Strategy is the same for most of the calculations. If we want to prove an equation we express its LHS in terms of c/a operators, shuffle them using (\ref{aux}) and recombine them to obtain the RHS. This procedure is often rather straightforward, but sometimes involves a tricky step or two. Writing down everything would be overwhelming (not to mention unnecessary), therefore we gather only the crucial steps here. 

As was mentioned, the important novelty for generalized $\kappa \neq 0$ states is that the left and the right multiplication by $r$ are unequal
\begin{eqnarray} \label{rLrR}
\hat{r}_L &=& \hat{r}+\rho, \ \hat{r}_R = \hat{r}-\rho , \ \rho = \frac{\lambda \kappa}{2} , \\ \nn
\hat{r}&=&\frac{1}{2}\left(\hat{r}_L + \hat{r}_R\right), \\ \nn
  \lambda \kappa &=& \hat{r}_L -\hat{r}_R=2 \rho .
\end{eqnarray}

One needs to go through the same calculations as in \cite{vel,LRL}, identify where the assumption $\hat{r}_L=\hat{r}_R$ was used and track down the corrections using (\ref{rLrR}).

It is very useful to use auxiliary operators
\begin{eqnarray} \label{ab}
&\hat{a}_\alpha \psi = a_\alpha \psi , \, & \hat{a}_\alpha^+ \psi = a_\alpha^+ \psi , \\ \nonumber
&\hat{b}_\alpha \psi = \psi a_\alpha , \, & \hat{b}_\alpha^+ \psi = \psi a_\alpha^+ 
\end{eqnarray}
and their quadratic combinations
\begin{eqnarray}\label{auxALG}
\hat{w}_{\alpha \beta} = \hat{a}^+_\alpha \hat{b}_\beta -\hat{a}_\beta \hat{b}^+_\alpha ,\ && \hat{\zeta}_{\alpha \beta} = \hat{a}^+_\alpha \hat{b}_\beta + \hat{a}_\beta \hat{b}^+_\alpha   , \\ \
\hat{\chi}_{\alpha \beta} = \hat{a}^+_\alpha \hat{a}_\beta + \hat{b}_\beta \hat{b}^+_\alpha \nonumber, \ && \hat{\mathcal{L}}_{\alpha \beta} = \hat{a}^+_\alpha \hat{a}_\beta - \hat{b}_\beta \hat{b}^+_\alpha \nonumber .
\end{eqnarray}

Most of the physical operators can be expressed using those either after contracting the indices $\alpha, \beta$ together ($\hat{A}_{\alpha \alpha} = \hat{A}$) or with those of Pauli matrices ($\hat{A}_{\alpha \beta} \sigma^i_{\alpha \beta} = \hat{A}_i$). For example $\hat{L}_i =  \frac{1}{2}\hat{\mathcal{L}}_i$, $\hat{X}_i = \frac{\lambda}{2} \hat{\chi}_i$, $\hat{r} = \frac{\lambda}{2}( \hat{\chi}+2 )$, $\hat{V}_i = \frac{i}{2r} \hat{w}_i$, $\hat{H}_0= \frac{1}{2\lambda r}(\chi - \zeta + 2)$. \footnote{This is the reason behind the peculiar names of the auxiliary operators, they are closely related to objects that have already been defined.}

\textbf{The velocity commutator}\\
This calculation is almost a carbon copy of the one for $\kappa = 0$ states in \cite{LRL}, the only modification appears right before the final step 
\begin{eqnarray} 
\varepsilon_{ijk} [ \hat{V}_i, \hat{V}_j] &=& \left( \mbox{same steps as for $\kappa=0$ states} \right)\\ \nn
&=&\frac{-\frac{i}{2}\sigma^k_{\alpha \delta}}{\hat{r}^2} ( \frac{\lambda}{\hat{r}} (\cancel{\hat{a}^+_\alpha \hat{b}_\beta \hat{a}^+_\beta \hat{b}_\delta} + \hat{a}_\beta \hat{b}^+_\alpha \hat{a}^+_\beta \hat{b}_\delta - \cancel{\hat{a}_\beta \hat{b}_\alpha^+ \hat{a}_\delta \hat{b}^+_\beta} - \hat{a}^+_\alpha \hat{b}_\beta \hat{a}_\delta \hat{b}^+_\beta \\ \nn
&& - \cancel{\hat{a}^+_\beta \hat{b}_\delta \hat{a}^+_\alpha  \hat{b}_\beta} - \hat{a}_\delta \hat{b}^+_\beta \hat{a}^+_\alpha \hat{b}_\beta + \hat{a}^+_\beta \hat{b}_\delta \hat{a}_\beta \hat{b}^+_\alpha + \cancel{\hat{a}_\delta \hat{b}^+_\beta \hat{a}_\beta \hat{b}_\alpha ^+} )  \\ \nn
&&+(\cancel{\hat{a}_\alpha^+\hat{b}_\beta \hat{a}^+_\beta \hat{b}_\delta}-\hat{a}^+_\alpha \hat{b}_\beta \hat{a}_\delta \hat{b}^+_\beta - \hat{a}_\beta\hat{b}^+_\alpha \hat{a}^+_\beta \hat{b}_\delta + \cancel{\hat{a}_\beta \hat{b}^+_\alpha \hat{a}_\delta \hat{b}_\beta^+} \\ \nn
&& - \cancel{\hat{a}_\beta^+ \hat{b}_\delta \hat{a}_\alpha^+ \hat{b}_\beta} + \hat{a}_\delta \hat{b}^+_\beta \hat{a}^+_\alpha \hat{b}_\beta - \cancel{\hat{a}_\delta \hat{b}^+_\beta \hat{a}_\beta \hat{b}^+_\alpha} + \hat{a}_\beta^+ \hat{b}_\delta \hat{a}_\beta \hat{b}^+_\alpha )) \\ \nn
&=&\frac{-\frac{i}{2}\sigma^k_{\alpha \delta}}{\hat{r}^2} \left( \frac{\lambda}{\hat{r}}\left( \frac{2\hat{r}_L}{\lambda}\hat{b}^+_\alpha \hat{b}_\delta - \frac{2\hat{r}_R}{\lambda}\hat{a}^+_\alpha \hat{a}_\delta \right)+\hat{a}_\delta \hat{a}^+_\alpha [\hat{b}^+_\beta, \hat{b}_\beta] + \hat{b}_\delta \hat{b}^+_\alpha [\hat{a}^+_\beta , \hat{a}_\beta]\right) \\ \nn
&=&\frac{-i}{\hat{r}^2 - \lambda^2 }\left( \frac{1}{\hat{r}} \frac{\hat{r}_L \hat{X}_{R,k} - \hat{r}_R \hat{X}_{L,k}}{\lambda}+ \frac{\hat{X}_{L,k} - \hat{X}_{R,k}}{\lambda} \right) \\ \nn
&=&\frac{-i}{\hat{r}(\hat{r}^2 - \lambda^2)}\frac{2 \rho}{\lambda}  \hat{X}_k ,
\end{eqnarray}
which is equal to
\begin{equation}
[\hat{V}_i,\hat{V}_j] = \varepsilon_{ijk} \frac{-i \left(\frac{\kappa}{2}\right) \hat{X}_k}{\hat{r}(\hat{r}^2-\lambda^2)}.
\end{equation}

\textbf{Square of the velocity operator and the (free) Hamiltonian}\\
For this calculation, it is convenient to express the velocity operator using (\ref{auxALG}) (pairs of terms with contracted indices are put into parenthesis as $a^+_\alpha a_\alpha = (a^+a)$)
\begin{eqnarray}
\hat{V}_i \hat{V}_i &=& - \frac{1}{4 \hat{r}} \sigma^i _{\alpha \beta} \sigma^i _{\gamma \delta} \hat{w}_{\alpha \beta} \frac{1}{\hat{r}} \hat{w}_{\gamma \delta} \\ \nn
&=&-\frac{1}{4\hat{r}}\left( 2 \delta_{\alpha \delta} \delta_{\beta \gamma} - \delta_{\alpha \beta} \delta_{\gamma \delta} \right) \left( \left(\frac{1}{\hat{r}-\lambda}\hat{a}^+_\alpha \hat{b}_\beta - \frac{1}{\hat{r}+\lambda}\hat{a}_\beta \hat{b}^+_\alpha \right)\left( \hat{a}^+_\gamma \hat{b}_\delta - \hat{a}_\delta \hat{b}^+_\gamma \right) \right) \\ \nn
&=&-\frac{1}{4\hat{r}}  \frac{1}{\hat{r}-\lambda}\left( 2\hat{a}^+_\alpha \hat{b}_\beta (\hat{a}^+_\beta \hat{b}_\alpha - \hat{a}_\alpha \hat{b}^+_\beta) - \hat{a}^+_\alpha \hat{b}_\alpha (\hat{a}^+_\delta \hat{b}_\delta - \hat{a}_\delta \hat{b}^+_\delta)\right) \\ \nn
&& +\frac{1}{4\hat{r}}  \frac{1}{\hat{r}+\lambda}\left(2 \hat{a}_\beta \hat{b}^+_\alpha (\hat{a}^+_\beta \hat{b}_\alpha - \hat{a}_\alpha \hat{b}^+_\beta) - \hat{a}_\alpha \hat{b}^+_\alpha (\hat{a}^+_\delta \hat{b}_\delta - \hat{a}_\delta \hat{b}^+_\delta)\right) \\ \nn
&=&-\frac{1}{4\hat{r}}  \frac{1}{\hat{r}-\lambda}\left( \cancel{2} (\hat{a}^+\hat{b})^2 - 2(\hat{a}^+\hat{a})(\hat{b}\hat{b}^+) - \cancel{(\hat{a}^+\hat{b})^2} +(\hat{a}^+\hat{b})(\hat{a}\hat{b}^+) \right) \\ \nn
&&+\frac{1}{4\hat{r}}  \frac{1}{\hat{r}+\lambda} \left( 2 (\hat{a}\hat{a}^+)(\hat{b}^+\hat{b}) - (\hat{a}\hat{b}^+)^2 - (\hat{a}\hat{b}^+)(\hat{a}^+\hat{b}) + \cancel{(\hat{a}\hat{b}^+)^2}\right) \\ \nn
&=& - \frac{1}{4\hat{r}} \frac{1}{\hat{r}-\lambda} \left((\hat{a}^+\hat{b})^2 - 2 \frac{\hat{r}_L-\lambda}{\lambda}\frac{\hat{r}_R-\lambda}{\lambda}+(\hat{a}^+\hat{b})(\hat{a}\hat{b}^+)\right) \\ \nn
&& -\frac{1}{4\hat{r}}  \frac{1}{\hat{r}+\lambda}\left( (\hat{a}\hat{b}^+)^2 - 2 \frac{\hat{r}_L+\lambda}{\lambda}\frac{\hat{r}_R+\lambda}{\lambda} +(\hat{a}\hat{b}^+)(\hat{a}^+\hat{b})\right) \\ \nn
&=& - \frac{1}{4\hat{r}} \left( \frac{1}{\hat{r}-\lambda} \left((\hat{a}^+\hat{b})^2 + (\hat{a}^+\hat{b})(\hat{a}\hat{b}^+)\right) +\frac{1}{\hat{r}+\lambda}\left( (\hat{a}\hat{b}^+)^2 + (\hat{a}\hat{b}^+)(\hat{a}^+\hat{b})\right)\right) \\ \nn
&& + \frac{1}{2\hat{r}\lambda^2} \left( \frac{(\hat{r}-\lambda+\rho)(\hat{r}-\lambda - \rho)}{\hat{r}-\lambda} + \frac{(\hat{r} + \lambda+\rho)(\hat{r}+\lambda-\rho )}{\hat{r}+\lambda} \right)\\ \nn
&=& - \frac{1}{4\hat{r}}\left( \frac{1}{\hat{r}-\lambda}\left((\hat{a}^+\hat{b})^2 + (\hat{a}^+\hat{b})(\hat{a}\hat{b}^+)\right) + \frac{1}{\hat{r}+\lambda}\left((\hat{a}\hat{b}^+)^2 + (\hat{a}\hat{b}^+)(\hat{a}^+\hat{b})\right)\right) \\ \nn
&&+\frac{1}{\cancel{2\hat{r}}\lambda^2}\cancel{2\hat{r}} \left( 1 - \frac{\rho^2}{\hat{r}^2 - \lambda^2}\right) .
\end{eqnarray}
To identify the $(ab)$ terms we first take
\begin{equation}
\hat{H}_0 - \frac{1}{\lambda^2} = - \frac{1}{2\lambda \hat{r}} \left((\hat{a}^+\hat{b}) + (\hat{b}^+\hat{a})\right),
\end{equation}
and square it to
\begin{eqnarray}
\left(\hat{H}_0 - \frac{1}{\lambda^2} \right)^2 &=& \frac{1}{2\lambda \hat{r}}  \left((\hat{a}^+\hat{b}) + (\hat{b}^+\hat{a})\right) \frac{1}{2\lambda \hat{r}}  \left((\hat{a}^+\hat{b}) + (\hat{b}^+\hat{a})\right)  \\ \nn
&=& \frac{1}{4 \lambda^2 \hat{r}} \left( \frac{1}{\hat{r}-\lambda}\left( (\hat{a}^+\hat{b})^2 +(\hat{a}^+\hat{b})(\hat{b}^+\hat{a})\right) +\frac{1}{\hat{r}+\lambda}\left( (\hat{b}^+\hat{a})^2 + (\hat{b}^+\hat{a})(\hat{a}^+\hat{b})\right)\right).
\end{eqnarray}
Comparing these two expressions we obtain
\begin{equation}
\lambda^2 \left( \hat{H}_0 - \frac{1}{\lambda^2}\right)^2 = - \hat{V}^2 + \frac{1}{\lambda^2}\left(1- \frac{\rho^2}{\hat{r}^2-\lambda^2}\right) ,
\end{equation}
or equivalently
\begin{equation}
\hat{V}_a^2 = \frac{1}{\lambda^2}\left( 1 -\frac{\rho^2}{\hat{r}^2-\lambda^2} \right),
\end{equation}
where $a=1,...,4$ (recall that $\hat{V}_4 = \frac{1}{\lambda}- \lambda \hat{H}_0$).

\textbf{The Coulomb problem}\\
Derivation of the Coulomb system spectrum in an algebraic way (developed by Pauli) is done in detail in \cite{LRL}. There, it was first shown that the Laplace-Runge-Lenz (LRL) vector defined as $\hat{A}_k =\frac{1}{2}\varepsilon_{ijk} (\hat{L}_i \hat{V}_j + \hat{V}_j\hat{L}_i) + q \frac{\hat{X}_k}{\hat{r}} $ can be expressed using (\ref{auxALG}) as  $\hat{A}_k = -\frac{1}{2\lambda \hat{r}}(\hat{r}\hat{\zeta} _k - \hat{X}_k \hat{\zeta}) + q \frac{\hat{X}_k}{\hat{r}}$. The Schr\"odinger equation can be, after restricting on energy eigenstates, expressed as $\hat{W}' = 2 \lambda q$, where $\hat{W}'= \eta \hat{r} - \hat{\zeta}$ with $\eta =  \frac{2}{\lambda} + \omega $ and $ \omega =  -2 \lambda E$.

Afterwards, it is shown that the LRL vector together with the angular momentum operator satisfy 
 \begin{equation}
[\hat{A}_i,\hat{A}_j]  =\frac{1}{4\lambda^2} [\hat{W}'_i, \hat{W}'_j]  = i\frac{\omega}{\lambda}\left(1+\frac{\omega\lambda}{4}\right)\varepsilon_{ijk} \hat{L}_k =i \varepsilon_{ijk} \left(- 2E + \lambda^2 E^2\right) \hat{L}_k ,
\end{equation}
\begin{equation}\label{LA}
[\hat{L}_i, \hat{L}_j] = i \varepsilon_{ijk} \hat{L}_k, \ [\hat{L}_i,\hat{A}_j] = i \varepsilon_{ijk} \hat{A}_k ,\  [\hat{L}_i, \hat{H}] = [\hat{A}_i, \hat{H}] =0  .
\end{equation}
Perhaps rather surprisingly this is not affected by considering $\kappa \neq 0$ states at all. The only differences appear for the Casimir operators, which are used to derive the energy spectrum. The first Casimir operator follows easily from
\begin{eqnarray}
\hat{X}_i \hat{L}_i &=& \frac{1}{4\lambda}(\hat{X}_{L,i} + \hat{X}_{R,i})(\hat{X}_{L,i} - \hat{X}_{R,i}) = \frac{1}{4\lambda}(\hat{X}_L^2 - \hat{X}_R^2) =  \frac{1}{4\lambda}(\hat{r}_L^2 - \hat{r}_R^2)  \\ \nn
&=&\frac{1}{4\lambda}(\hat{r}_L+\hat{r}_R)(\hat{r}_L-\hat{r}_R) = \frac{1}{2\lambda}\hat{r} (\hat{r}_L-\hat{r}_R) =  \frac{\kappa}{2}\hat{r} , \\ \nn
\hat{L}_j\hat{\zeta}_j &=& \frac{1}{2\lambda}\left((\hat{r}_L-\hat{r}_R)\hat{a}^+_\alpha \hat{b}_\alpha + (\hat{r}_L -\hat{r}_R) a_\alpha \hat{b}^+_\alpha \right) =  \frac{\kappa}{2}\hat{\zeta}, 
\end{eqnarray}
as
\begin{equation}
\hat{C}_1' = \hat{L}_j \hat{A}_j = \frac{1}{2\lambda} \hat{L}_j (\eta \hat{X}_j - \hat{\zeta}_j)= -\frac{1}{2\lambda}\left(-\frac{\kappa}{2}\right) (\eta \hat{r} - \hat{\zeta}) = \frac{\kappa}{2}q .
\end{equation}

The $\kappa \neq 0$ correction is apparent. Derivation of the second Casimir operator is considerably more complicated, the RHS of the following equation is a constant and we need to identify its value
\begin{equation} \label{Peq}
\hat{C}_2' = \hat{W}'_i \hat{W}'_i +(\eta^2 \lambda^2 - 4)(\hat{L}_i \hat{L}_i+1)  \, .
\end{equation}

Expressing the terms on the RHS we obtain (after a number of auxiliary calculations)
\begin{eqnarray}
 \hat{W}'_i \hat{W}'_i &+&(\eta^2 \lambda^2 - 4)(\hat{L}_i \hat{L}_i+1) \\ \nn
 &=& \eta^2 \hat{X}^2 - \eta\{\hat{X}_i, \hat{\zeta}_i \} + \hat{\zeta}^2 +\frac{2}{\lambda^2}\left( \hat{r}_L \hat{r}_R - \hat{X}_{L,i} \hat{X}_{R,i} + \lambda^2 \right)  \\ \nn
 &&+ \eta^2 \frac{1}{4}(\hat{X}_L^2 +\hat{X}_R^2 - 2 \hat{X}_{L,i} \hat{X}_{R,i} )- \frac{1}{\lambda^2}(\hat{X}_L^2 + \hat{X}_R^2 - 2 \hat{X}_{L,i} \hat{X}_{R,i}) + \eta^2 \lambda^2 - 4 \\ \nn
 &=&\left( - \eta\{\hat{r},\hat{\zeta}\}  + \hat{\zeta}^2 \right) +\frac{\eta^2}{4}\left(\hat{X}_L^2 +\hat{X}_R^2 +\cancel{2 \hat{X}_{L,i} \hat{X}_{L,i}} \right)+ \frac{2}{\lambda^2}(\hat{r}_L\hat{r}_R - \cancel{\hat{X}_{L,i}\hat{X}_{R,i}} + \lambda^2) \\ \nn
 &&+ \eta^2 \frac{1}{4}(\hat{X}_L^2 +\hat{x}_R^2 - \cancel{2 \hat{X}^L_i \hat{X}^R_i}) - \frac{1}{\lambda^2} (\hat{X}_L^2 + \hat{X}_R^2 - \cancel{2 \hat{X}_{L,i} \hat{X}_{R,i}}) + \eta^2 \lambda^2 - 4 \\ \nn
 &=&\left( - \eta\{\hat{r},\hat{\zeta}\}  + \hat{\zeta}^2 \right) + \frac{\eta^2}{\cancel{2}}\left( \cancel{2} (\hat{r}^2 - \cancel{\lambda^2}) + \cancel{2} \lambda^2 \left(\frac{\kappa}{2}\right)^2\right) - \frac{1}{\lambda^2}\left( 2 (\cancel{\hat{r}^2} - \cancel{\lambda^2}) + 2 \lambda^2 \left(\frac{\kappa}{2}\right)^2 \right) \\ \nn
 &&+\frac{2}{\lambda^2}\left(\cancel{\hat{r}^2} - \lambda^2  \left(\frac{\kappa}{2}\right)^2 + \cancel{\lambda^2}\right) + \cancel{\eta^2 \lambda^2} - \cancel{4} \\ \nn
 &=& (\hat{W}')^2 + \eta^2 \lambda^2  \left(\frac{\kappa}{2}\right)^2 - 2  \left(\frac{\kappa}{2}\right)^2 - 2  \left(\frac{\kappa}{2}\right)^2 = 4 \lambda^2 q^2 + (\eta^2 \lambda^2 - 4)  \left(\frac{\kappa}{2}\right)^2 .
\end{eqnarray}
\begin{acknowledgments}
This work was partially supported by COST action MP1405 (QSPACE) and by VEGA project 1/0985/16. 
\end{acknowledgments}

\bibliography{your-bib-file}

\end{document}